\documentclass[12pt]{article}
\usepackage{graphicx,psfrag,epsf}
\usepackage{enumerate}
\usepackage{natbib}
\usepackage{url} 
\usepackage{hyperref}
\usepackage{amsmath,amssymb,amsfonts,bm,bbm,amsthm} 
\usepackage{color}
\usepackage{enumitem}
\usepackage{mathtools}
\usepackage{algpseudocode}
\usepackage{algorithmicx, algorithm}
\usepackage{booktabs}
\usepackage{subcaption}
\usepackage{xr}
\usepackage[T1]{fontenc}
\usepackage[utf8]{inputenc}

\numberwithin{equation}{section}
\newtheorem{theorem}{Theorem}[section]

\def\HH{\mathcal H}

\def\R{\mathbb R}

\newcommand{\blind}{1}

\begin{document}

\def\spacingset#1{\renewcommand{\baselinestretch}%
{#1}\small\normalsize} \spacingset{1}


\if1\blind
{
  \title{\bf High-Dimensional Smoothing Splines and
Application in Alzheimer’s Disease Prediction
Using Magnetic Resonance Imaging}
  \author{Xiaowu Dai\thanks{ University of California, Berkeley, CA (E-mail: \href{mailto:xwdai@berkeley.edu}{xwdai@berkeley.edu})}  
 } 
    \date{}
  \maketitle
} \fi

\if0\blind
{
  \bigskip
  \bigskip
  \bigskip
  \begin{center}
    {\Large \bf Alzheimer's Disease Prediction  \\ Using Longitudinal and Heterogeneous Magnetic Resonance  Imaging}
\end{center}
  \medskip
} \fi


\begin{abstract}
Recent evidence has shown that structural magnetic resonance imaging (MRI) is an effective tool for Alzheimer’s disease (AD) prediction. While traditional MRI-based prediction uses images acquired at a single time point, a longitudinal study is more sensitive and accurate in detecting early pathological changes of the AD. Two main statistical difficulties arise in the longitudinal MRI-based analysis: (i) the inconsistent longitudinal scans among subjects (i.e., the different scanning time and the different total number of scans); (ii) the heterogeneous progressions of high-dimensional regions of interest (ROIs) in MRI. In this work, we propose a new feature selection and estimation method which can be applied to extract AD-related features from the heterogeneous longitudinal MRI.  A key ingredient of our approach is a hybrid of the smoothing splines and the $l_1$-penalty. Smoothing splines can integrate information from heterogeneous progressions of ROIs and adapt to inconsistent scans of MRIs. The selection property of the $l_1$-penalty helps to select important ROIs related to AD. We introduce an efficient algorithm to perform the proposed method. Real data experiments on the Alzheimer’s Disease Neuroimaging Initiative database are provided to corroborate some advantages of the proposed method for AD prediction in longitudinal studies. 
\end{abstract}

\noindent%
\vfill

\newpage
\spacingset{1.45} 

\section{Introduction}
\label{sec:intro}

Alzheimer’s Disease (AD) is the most common cause of dementia in the aged population  \citep{prince2013global}.
It is vital to identify AD-related pathological biomarkers and diagnose early-stage AD to prevent disease progression and take treatment in the earliest stage. A considerable amount of research has been devoted to the use of structured magnetic resonance imaging (MRI) for early-stage AD diagnosis; e.g., \citet{jack2010hypothetical, jack2013tracking}. The structural MRI provides measures of cerebral atrophy, and it is shown to be closely coupled with clinical symptoms in AD  \citep{jack2009serial}.

\begin{figure}[htp]
\centering
\includegraphics[width=0.85\textwidth]{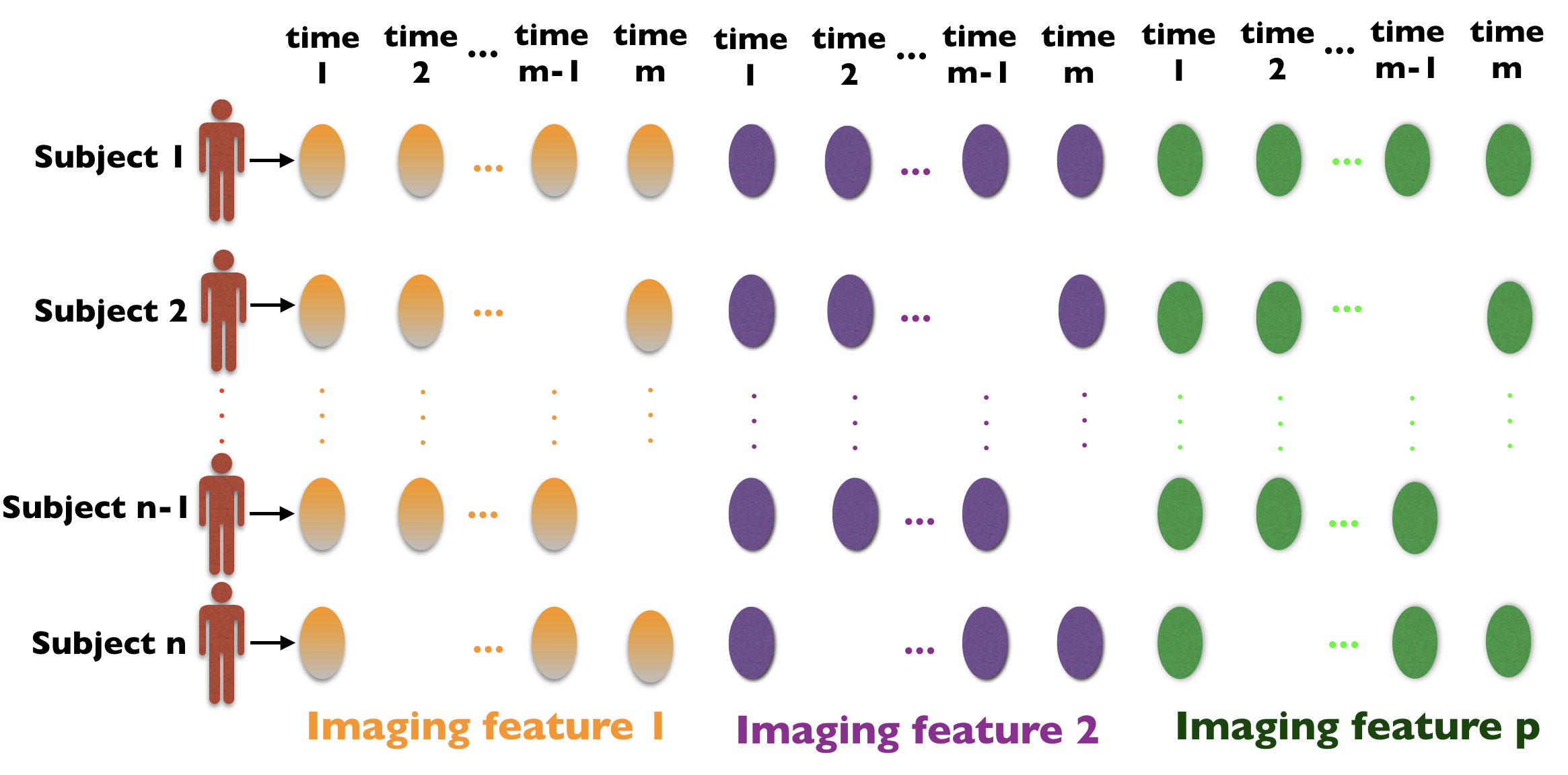}
\caption{Illustration of heterogeneous longitudinal data with $n$ subjects, $p$ features, and $m$ time points.}
\label{longitudinal}
\end{figure}

Most works consider the cross-sectional study with MRI acquired at one single time point; see, e.g., \citet{tzourio2002automated}, \citet{aguilar2013different}, and  \cite{liu2016relationship}.
However, the cross-sectional study could be insensitive  to early pathological changes. As an alternative, longitudinal analysis of structural abnormalities has recently attracted  attention \citep{zhang2012predicting, yau2015longitudinal,  chincarini2016integrating}. Existing longitudinal studies  focus on the atrophy of a few well-known regions of interest (ROIs) such as the hippocampus, entorhinal cortex, and ventricular cortex. However, the prespecified ROIs are insufficient to capture the full morphological abnormality pattern of the brain MRI. A few other issues also remain as challenges in the longitudinal analysis. First,  longitudinal scans across subjects are usually inconsistent. For example, subjects could have different scanning time and the  different total number of scans. Second, the total number of  ROIs  in the brain is enormous compared with the number of subjects, which poses a challenge to select AD-related longitudinal biomarkers from the whole brain.  Third, the rates of longitudinal change of ROIs are different, and this heterogeneity has not been considered in the modeling of AD progression.

Our goal is to identify AD-related ROIs in the whole brain with longitudinal MRI and perform AD prediction using selected ROIs. Specifically,
we consider the varying coefficient model \citep{hastie1993varying} to characterize the heterogeneous changes of ROIs in the structural MRI. This model allows the nonlinear functional modeling between MRI and clinical cognition functions. 
We propose a novel feature selection method for high-dimensional varying coefficient models, where the key idea is combining  the smoothing splines \citep{wahba1990spline} and an $l_1$-penalty  \citep{tibshirani1996regression}. Our method can simultaneously select and estimate 
AD-related  ROIs. 
We also provide an efficient algorithm to implement the 
proposed feature selection method.
Then the prediction is performed based on the
 selected longitudinal features and the estimated varying coefficients. 
Our approach is robust to the inconsistency among longitudinal scans and is adaptive to the heterogeneity of changes in different ROIs.
The hypothetical AD dynamic biomarker curves \citep{jack2010hypothetical, jack2013tracking} motivate the use of varying coefficient models in our approach. In particular, \citet{jack2010hypothetical} suggests that  the rates of change over time for MRI and clinical cognition functions are in a temporally ordered manner, which implies the functional relationship between the atrophy of MRI  and the change of cognition functions is nonlinear with time.

To evaluate our method, we conduct  experiments using data from the  Alzheimer's Disease Neuroimaging Initiative (ADNI). We predict future clinical changes of mild cognitive impairment (MCI) subjects with longitudinal  MRI data. The MCI is a prodromal stage of AD and the prediction of clinical changes helps to determine whether an MCI subject will convert into AD at a future time point, which is vital for early diagnosis of AD.

Main differences between this paper and existing longitudinal studies in \citet{zhang2012predicting}, \citet{yau2015longitudinal}, and \cite{chincarini2016integrating} are listed as follows.
\begin{itemize}
\item Different feature representations. We consider the varying coefficient model to  characterize the nonlinear and smooth progression of longitudinal features, which is motivated by clinical findings and dynamic biomarker curves in  \cite{jack2010hypothetical, jack2013tracking}.
On the contrary,  \citet{zhang2012predicting}, \citet{yau2015longitudinal}, and \cite{chincarini2016integrating} consider only linear representations for features. 
\item Different scalability to heterogeneous longitudinal scans. Our method does not require identical scanning times or an equal number of scans across samples. However, these conditions are necessary for  \citet{zhang2012predicting}, \citet{yau2015longitudinal}, and \cite{chincarini2016integrating}.
\item Different feature selection methods. We proposed a new feature selection method by combining the smoothing splines with an $l_1$-penalty, which enables simultaneous  feature selection and varying coefficient estimation. 
The method is different from  \cite{zhang2012predicting},  which performs the feature selection and estimation separately in a two-step procedure,  and it is also different from \citet{ yau2015longitudinal} and \citet{chincarini2016integrating}, which only use prespecified features.
\end{itemize}

The rest of the paper is organized as follows. We introduce our method  in Section \ref{sec:materialmethod}. 
We give the experiment results in Section \ref{sec:experiment}.
The concluding remarks and discussions are provided in Section \ref{sec:discussion}.
Additional material and proofs are relegated to the Appendix.

\section{Methodology}
\label{sec:materialmethod}

The varying coefficient model describes time-dependent covariate effects on the response  \citep{hastie1993varying}. 
Given the scaled time $t\in[0,1]$, the response functional $Y(\cdot)$ is related to 
covariates
 $X_1(\cdot), \ldots, X_p(\cdot)$ through   
\begin{equation}
\label{eqn:varymodel}
Y(t) = b+\sum_{j=1}^p\beta_j(t)X_j(t)  + \varepsilon(t),\quad b\in\R,
\end{equation}
where the centered noise process $\varepsilon(\cdot)$ is independent of covariates $X_j(\cdot)$s.
The model (\ref{eqn:varymodel}) allows a nonlinear functional relationship between $X_j(\cdot)$s and $Y(\cdot)$ as the coefficients $\beta_j(\cdot)$s vary over $t\in[0,1]$. 
Take Alzheimer's disease prediction in Section \ref{sec:experiment} as an example. The response represents the clinical cognitive test score, and the covariates include demographic information (such as age, gender, and education years) and ROIs in the structural brain MRI.
The dependence of $\beta_j(\cdot)$s on $t$ implies the time-varying effects of covariates  on the response.
On the other hand, model (\ref{eqn:varymodel}) has an additive structure on covariates $X_j(\cdot)$s to allow efficient estimation of the coefficients $\beta_j(\cdot)$s.

In practice, we observe data for subjects $i=1,\ldots,n$ at time $t_{i\nu}$, where $\nu=1,2,\ldots,m_i$ and $0\leq t_{i1}\leq t_{i2}\leq \cdots\leq t_{im_i}\leq1$.
Here, $m_i$ and $t_{i\nu}$s can be different for different subjects. 
Let $x_{ij}(\cdot)$ be the observation of covariate $X_j(\cdot)$ for subject $i$. Let
$y_{i\nu}$ be the   response for the subject $i$ at time $t_{i\nu}$. Then model (\ref{eqn:varymodel}) suggests
\begin{equation}
 \label{VCmodel}
 y_{i\nu} = b+ \sum_{j=1}^p\beta_j(t_{i\nu})x_{ij}(t_{i\nu}) + \varepsilon(t_{i\nu}),\quad b\in\R.
 \end{equation} 
 The structure of heterogeneous longitudinal data is illustrated in Figure \ref{longitudinal}, where some subjects could have missing  feature values at certain time points.
The number of covariates $p$ in (\ref{VCmodel}) can be larger than the sample size $n$, where (\ref{VCmodel}) becomes a high-dimensional model. 
Since some covariates can be irrelevant with the response, we want to select  relevant covariates $X_j(\cdot)$s based on data (\ref{VCmodel}). Then we use the chosen covariates for prediction.

 We  propose a new method to simultaneously select covariates and estimate their  varying coefficients as follows. 
 Assume that varying coefficients $\beta_1(\cdot),\beta_2(\cdot),\ldots,\beta_p(\cdot)$ reside in a  reproducing kernel Hilbert space (RKHS), $(\HH_K,\|\cdot\|_{\HH_K})$, where the reproducing kernel is denoted by $K(\cdot,\cdot)$ \citep{wahba1990spline}. We want to find $\beta_1(\cdot),\beta_2(\cdot),\ldots,\beta_p(\cdot)\in\HH_K$ and $b\in\R$ to minimize
 \begin{equation}
\label{eqn:highdimenscheme}
\begin{aligned}
 &  \frac{1}{N}\sum_{i=1}^n\sum_{\nu=1}^{m_i}\left[y_{i\nu} - b- \sum_{j=1}^p\beta_j(t_{i\nu})x_{ij}(t_{i\nu})\right]^2 + \lambda\sum_{j=1}^p\|\beta_j\|_{\HH_K},
 \end{aligned}
\end{equation}
where $N=\sum_{i=1}^nm_i$ and $\|\cdot\|_{\HH_K}$ denotes the  RKHS norm. 
Note that the measurements for the same  covariate at different time points:
  $x_{ij}(t_{i1}),\ldots,x_{ij}(t_{im_i})$, are highly correlated since they are observed from the same
random function $x_{ij}(\cdot)$, $j=1,\ldots,p$. The proposed method (\ref{eqn:highdimenscheme})  allows such correlation among measurements for the same covariate. 
The first term in (\ref{eqn:highdimenscheme}) measures the goodness of data fitting and the second term merits the selection property by the $l_1$-like penalty $\sum_{j=1}^p\|\beta_j\|_{\HH_K}$.
We first provide the following theorem to justify the existence of minimizer for (\ref{eqn:highdimenscheme}).
\begin{theorem}
\label{thm:existminimizer}
There exists a minimizer of (\ref{eqn:highdimenscheme}) that is in the domain $\beta_1(\cdot),\ldots, \beta_p(\cdot)\in\HH_K$ and $b\in\R$.
\end{theorem}
We give the proof of this theorem in Appendix \ref{proofthm:existminimizer}. 
The variable selection method (\ref{eqn:highdimenscheme}) is new in the literature, and it  is efficient for optimization due to the convexity in $\beta_j(\cdot)$s and having only one tuning parameter $\lambda$.
We provide an algorithm for giving a minimizer of (\ref{eqn:highdimenscheme})  in Appendix \ref{subsec:algorithm}.

The following theorem gives further insights into (\ref{eqn:highdimenscheme}), which is actually a combination of the smoothing splines \citep{wahba1990spline} and the Lasso \citep{tibshirani1996regression}.
\begin{theorem}
\label{thm:equivalence}
Consider the following optimization problem. Find $\beta_1(\cdot),\ldots,\beta_p(\cdot)\in\HH_K$ and $\theta_1,\ldots,\theta_p, b\in\R$ to minimize
\begin{equation}
\label{eqn:equivform}
 \begin{aligned}
   \frac{1}{N}\sum_{i=1}^n\sum_{\nu=1}^{m_i}\left[y_{i\nu} - b-\sum_{j=1}^p \beta_j(t_{i\nu})x_{ij}(t_{i\nu})\right]^2&+ \tau_0\sum_{j=0}^p \theta^{-1}_j\|\beta_j\|_{\HH_K}^2 + \tau_1\sum_{j=0}^p\theta_j,\\
 &\text{s.t. } \theta_j\geq 0, j=0,1,\ldots,p,
  \end{aligned}
\end{equation}
where $\tau_0$ is a constant  and $\tau_1$ is a tuning parameter. 
Let $\tau_1=\lambda^4/(4\tau_0)$. The following equivalence holds.
\begin{itemize}
\item[1)] If $\left(\widehat{\beta}_0, \widehat{\beta}_1(\cdot),\ldots,\widehat{\beta}_p(\cdot)\right)$ minimizes (\ref{eqn:highdimenscheme}), by letting $\widehat{\theta}_j = \tau_0^{1/2}\tau_1^{-1/2}\|\widehat{\beta}_j\|_{\HH_K}$, we have that $(\widehat{\theta}_1,\ldots,\widehat{\theta}_p; \widehat{\beta}_0, \widehat{\beta}_1(\cdot),\ldots,\widehat{\beta}_p(\cdot))$ minimizes (\ref{eqn:equivform}).
\item[2)] If there exists  $(\widehat{\theta}_1,\ldots,\widehat{\theta}_p; \widehat{\beta}_0, \widehat{\beta}_1(\cdot),\ldots,\widehat{\beta}_p(\cdot))$ minimizes (\ref{eqn:equivform}), then $(\widehat{\beta}_0, \widehat{\beta}_1(\cdot),\ldots,\widehat{\beta}_p(\cdot))$ minimizes (\ref{eqn:highdimenscheme}).
\end{itemize}
\end{theorem}
We give the proof of this theorem in Appendix \ref{proofofequivalence}. The first two terms in (\ref{eqn:equivform}):
\begin{equation*}
\frac{1}{N}\sum_{i=1}^n\sum_{\nu=1}^{m_i}[y_{i\nu} - b-\sum_{j=1}^p \beta_j(t_{i\nu})x_{ij}(t_{i\nu})]^2+ \tau_0\sum_{j=0}^p \theta^{-1}_j\|\beta_j\|_{\HH_K}^2
\end{equation*}
amount to the smoothing splines in nonparametric statistics  \citep{wahba1990spline}, which integrate information from heterogeneous progressions of ROIs through the varying coefficients $\beta_j(\cdot)$s and adapt to inconsistent scans of MRIs with different $m_i$s. The last term in (\ref{eqn:equivform}):
$\tau_1\sum_{j=0}^p\theta_j$
is the same as the $l_1$ Lasso penalty \citep{tibshirani1996regression} with weights $\theta_j$s, which helps to select features.

Now we consider to use the selected covariates for prediction. 
Let  $X_{j_1}, X_{j_2},\ldots, X_{j_s}$ be $s$ of selected features by (\ref{eqn:highdimenscheme}), where $1\leq j_1\leq j_2\leq \cdots\leq j_s\leq p$. Let $\widehat{\beta}_{j_1}(t), \widehat{\beta}_{j_2}(t),\ldots, \widehat{\beta}_{j_s}(t)$  be the corresponding  varying coefficients estimated by (\ref{eqn:highdimenscheme}). The prediction  for a new subject with features $X^*_{j_1}(t), X^*_{j_2}(t),\ldots, X^*_{j_s}(t)$ at time $t$ is given by
\begin{equation*}
\widehat{Y}^*(t) = \widehat{\beta}_{j_1}(t)X^*_{j_1}(t) + \widehat{\beta}_{j_2}(t)X^*_{j_2}(t)+ \cdots+\widehat{\beta}_{j_s}(t)X^*_{j_s}(t).
\end{equation*}
Note that the prediction for  subject $i=1,\ldots,n$ of the model (\ref{VCmodel}) at a future time point $t$ can also be  performed. Here, the selected features $x_{ij_1}(t),x_{ij_2}(t),\ldots,x_{ij_s}(t)$ of subject $i$ in the future time $t$ are usually unknown. 
There are two methods to estimate the values of $x_{ij_1}(t),x_{ij_2}(t),\ldots,x_{ij_s}(t)$. 
The first method is to apply nonparametric statistical approaches such as the smoothing splines \citep{wahba1990spline}. Assume that the covariates $X_j(\cdot)$s are smooth and belong to the RKHS $(\HH_K,\|\cdot\|_{\HH_K})$. We find $z(\cdot)\in\HH_K$ to minimize
\begin{equation*}
\frac{1}{m_i}\sum_{\nu=1}^{m_i}\left[z(t_{i\nu}) - x_{ij_1}(t_{i\nu})\right]^2+\lambda_1\|z\|_{\HH_K}^2,
\end{equation*}
where the smoothing parameter $\lambda$ can be chosen by  generalized cross-validation (GCV)  \citep{wahba1990spline}.
The feature $x_{ij_1}(t)$ at time $t$ is then estimated by evaluating $z(\cdot)$ at $t$. That is, $\widehat{x}_{ij_1}(t) = z(t)$.
Similarly, the smoothing splines can also be used to estimate other features: $x_{ij_2}(t),\ldots,x_{ij_s}(t)$. The second method to estimate the values of $x_{ij_1}(t),x_{ij_2}(t),\ldots,x_{ij_s}(t)$ is using the selected features  of subject $i$ at an observed time $\tilde{t}$ which is closest to time $t$: $\tilde{t}\in\{t_{i1},t_{i2},\ldots,t_{im_i}\}$ and $|\tilde{t}-t| = \min\{|t_{i_1}-t|,\ldots,|t_{im_i}-t|\}$. That is, $\widehat{x}_{ij_1}(t) = x_{ij_1}(\tilde{t}),\widehat{x}_{ij_2}(t) = x_{ij_2}(\tilde{t}),\ldots,\widehat{x}_{ij_s}(t) = x_{ij_s}(\tilde{t})$.
With the estimated covariates values $\widehat{x}_{ij_1}(t),\widehat{x}_{ij_2}(t),\ldots,\widehat{x}_{ij_s}(t)$, we give
the prediction for  subject $i$ in the future time $t$ by
\begin{equation*}
\widehat{y}_{i}(t) = \widehat{\beta}_{j_1}(t)\widehat{x}_{ij_1}(t) + \widehat{\beta}_{j_2}(t)\widehat{x}_{ij_2}(t)+ \cdots+\widehat{\beta}_{j_s}(t)\widehat{x}_{ij_s}(t).
\end{equation*}
The experiments in Section \ref{sec:experiment} have known features in the future time $t$ such as age, gender, and education years. However, the brain MRI data are unknown in future time. We 
consider the second method above  to estimate the values of covariates $x_{ij_1}(t),x_{ij_2}(t),\ldots,x_{ij_s}(t)$, that is, using the feature values of subject $i$ at an observed time $\tilde{t}$  which is closest to time $t$. The reasons are that the changes of MRI features are generally monotone and the
 variations of MRI features within subjects (i.e., longitudinal variations) are significantly smaller compared to that across subjects.

\section{Experiment Results}
\label{sec:experiment}

In this section, we predict future clinical changes of MCI subjects with real data from the ADNI database.
A detailed description of the ADNI database is relegated to Appendix \ref{sec:adni1}.
The MCI is a prodromal stage of AD. 
Generally, some MCI subjects will convert into AD after a certain time (i.e., MCI converters, MCI-C for short), while others will not convert (i.e., MCI non-converters, MCI-NC for short) \citep{zhang2012predicting}. 
The prediction of clinical change for an MCI subject helps to determine whether the subject will convert into AD at a future time, and it is a central task for the early diagnosis of AD. 
We summarize the baseline demographic information of ADNI subjects studied here  in Table \ref{table:demosubjects}.

\begin{table}[h]
\caption{Demographics  of ADNI subjects studied here.}
\center
	\resizebox{0.6\textwidth}{!}{
\begin{tabular}{@{\extracolsep{5pt}}   l c c } 
\toprule
 &  MCI-C ($n=74$)  &  MCI-NC ($n=98$) \\
			\midrule
Male $/$  Female & $44 \ / \ 30$ & $61 \ / \ 37$ \\ 
Age (years) & $73.03 \pm 6.65$ & $74.35 \pm 7.47$ \\ 
  Edu. (years) & $15.51 \pm 3.05$  & $15.59 \pm 3.07$\\
			\bottomrule
\end{tabular}
}
\label{table:demosubjects}
\end{table}

The preprocessing steps of brain MR imaging  are described in Appendix \ref{sec:preprocessbrainmri}. Specifically, we have a total of 324 ROIs for each imaging.
For MCI subjects, MRI scans were taken at baseline (bl), six months (M06), one year (M12), eighteen months (M18), two years (M24), three years (M36), and four years (M48). 
However, some subjects may miss a few visits, and they do not have MRI scans at these missed time points.
We choose $n=172$ MCI subjects who have M48 imaging data.
Table \ref{table:distofscans} lists the distributions of visit times for these $172$ MCI subjects, where, e.g.,  six of MCI-C subjects make at most three visits among the scheduled six times (bl, M06, M12, M18, M24, M36) such that they have at most three longitudinal MRI scans.
\begin{table}[h]
\caption{Distribution of visit times for ADNI subjects studied here.}
\center
	\resizebox{0.55\textwidth}{!}{
\begin{tabular}{@{\extracolsep{5pt}}  l c c } 
\toprule
 &  MCI-C ($n=74$)  &  MCI-NC ($n=98$) \\
			\midrule
$\leq3$ scans & $6$ & $6$ \\ 
$4$ scans & $8$  & $14$ \\ 
$5$ scans & $15$  & $33$\\
$6$ scans & $45$  & $45$\\ 
			\bottomrule
\end{tabular}
}
\label{table:distofscans}
\end{table}

Our goal is to use longitudinal information (from bl up to M36) to predict the clinical changes of MCI subjects at M48.
Empirical evidence suggests that the rates of change over time for structural MRI and clinical cognition functions are in a temporally ordered manner \citep{jack2010hypothetical, jack2013tracking}.
Hence, we consider the varying coefficient model (\ref{eqn:varymodel})  for the nonlinear modeling of the functional relationship between the atrophy of MRI and the change in clinical cognition functions.
The Alzheimer's Disease Assessment Scale--Cognitive Subscale (ADAS-Cog) is used as the
response clinical cognitive test score $Y(\cdot)$. The ADAS-Cog score ranges from 70 (severe cognitive impairment) to 0 (no cognitive impairment) and it measures disturbances of memory, language, and other cognitive abilities. 
The prediction of future clinical scores based on the information at previous time points  helps monitor the disease progression. 
We provide in Table \ref{table:adasinfo} the average ADAS-Cog scores at baseline and M48 time point together with the $p$-value for the difference. There exists a significant difference between  ADAS-Cog scores  of baseline and M48 for MCI-C group while no significant difference for MCI-NC group, which indicates that  ADAS-Cog scores for MCI-NC subjects increase much slower than that of the MCI-C subjects. In other words, 
Table \ref{table:adasinfo} suggests that  for an MCI subject if there is a significant increase in the prediction of ADAS-Cog score at M48 compared to ADAS-Cog score at baseline, the MCI subject is likely to be an MCI converter. 
Furthermore, a statistical classification model such as the support vector machines can also be built to classify MCI-C and MCI-NC subjects based on the predicted ADAS-Cog scores and selected features.

\begin{table}[h]
\caption{ADAS-Cog scores  of ADNI subjects studied here.}
\center
	\resizebox{0.65\textwidth}{!}{
\begin{tabular}{ @{\extracolsep{5pt}} l c c } 
\toprule
 &  MCI-C  ($n=74$) &  MCI-NC  ($n=98$) \\
			\midrule
ADAS-Cog (baseline) & $20.12\pm3.79$ & $16.01\pm3.91$  \\ 
ADAS-Cog (M48) & $25.73\pm4.22$ & $17.14\pm 4.16$ \\ 
$p$-value & $0.0031$  & $0.2188$\\
			\bottomrule
\end{tabular}
}
\label{table:adasinfo}
\end{table}

\begin{figure*}[h!]
\centering
\includegraphics[width=0.9\textwidth]{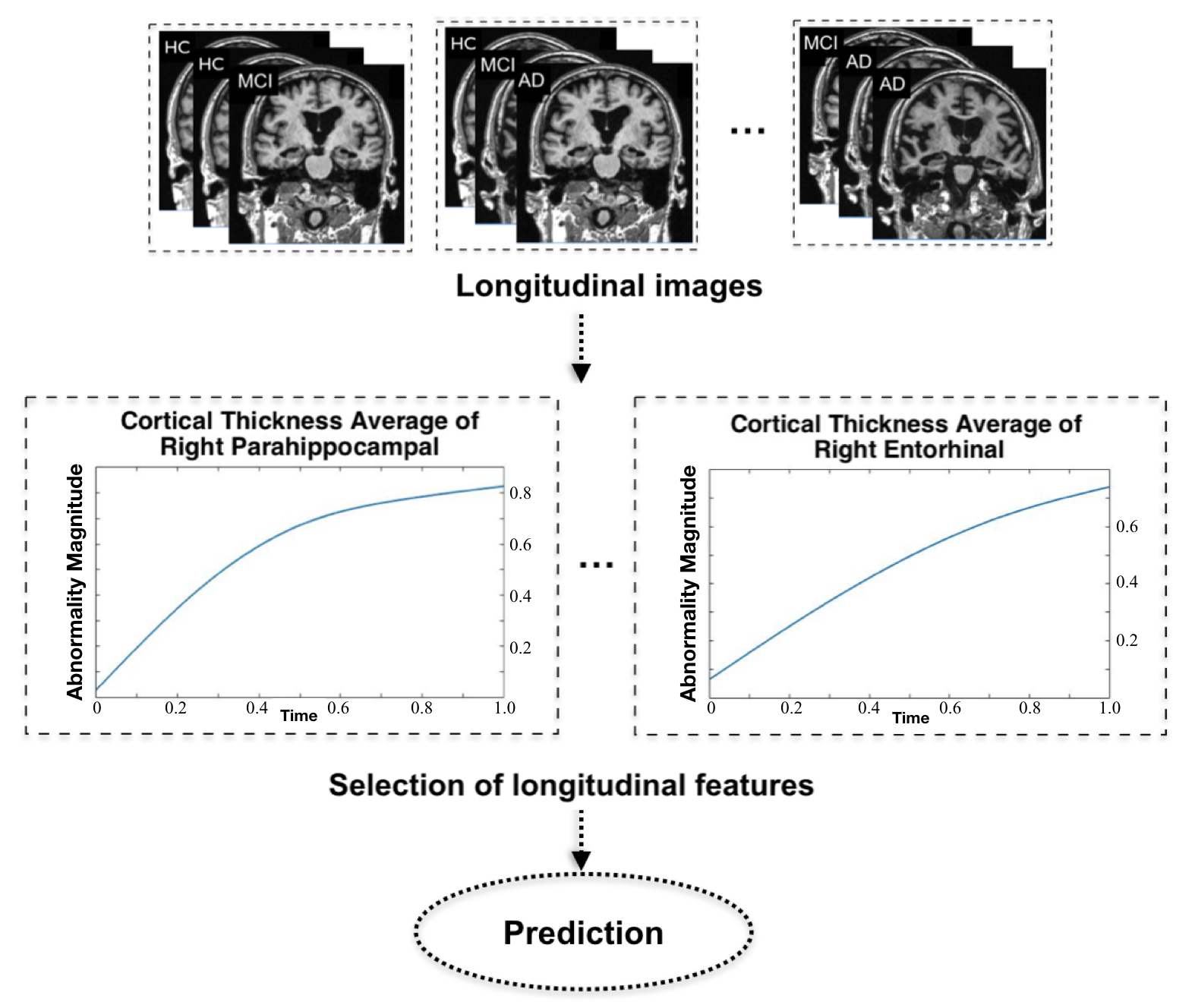}
\caption{Flowchart of the proposed method. In the middle panel, the $x$-axis ("Time") is scaled to $[0,1]$, where $t=0$ represents bl,  $t=1$ represents M48. The $y$-axis ("Abnormality Magnitude") is the quantity that measures the abnormal shrinkage of a feature,  that is, \\
$1- \text{(average shrinkage of a feature  of MCI-NC)}/\text{(average shrinkage of a feature  of MCI-C)}$.} 
\label{framework}
\end{figure*}

The covariates $X_j(\cdot)$s in the varying coefficient model  (\ref{eqn:varymodel})  consist of 324 MR imaging ROIs and three demographic covariates: age, gender, and education years. We let $t$ in (\ref{eqn:varymodel})  be
the scaled time relative to subjects entering the ADNI study. Hence, $t$ is identifiable.  
Figure \ref{framework} gives the flowchart of our approach, where the selected features (e.g., cortical thickness of the right parahippocampal cortex and  cortical thickness of the right entorhinal cortex) shown in the middle panel exhibit the abnormal shrinkages with time for the MCI-C. We build six models based on six different levels of longitudinal information in the training.
\begin{itemize}
\item Model 1 ($p=327,\max_{i} m_i=1$):  All subjects that have observations at bl.
\item Model 2 ($p=327,\max_{i} m_i=2$):  All subjects that have observations at bl or M06. 
\item Model 3 ($p=327,\max_{i} m_i=3$):  All subjects that have observations at bl or M06 or M12.
\item Model 4 ($p=327,\max_{i} m_i=4$):  All subjects that have observations at bl or M06 or M12 or M18.
\item Model 5 ($p=327,\max_{i} m_i=5$):  All subjects that have observations at bl or M06 or M12 or M18 or M24.
\item Model 6 ($p=327,\max_{i} m_i=6$):  All subjects that have observations at bl or M06 or M12 or M18 or M24 or M36.
\end{itemize}
In each of these six models, we allow the training subjects to have missing. For example, Model 3 includes subjects that only have observations at bl and M12 but missed M06.

\begin{figure*}[h!]
\centering
\includegraphics[width=0.72\textwidth]{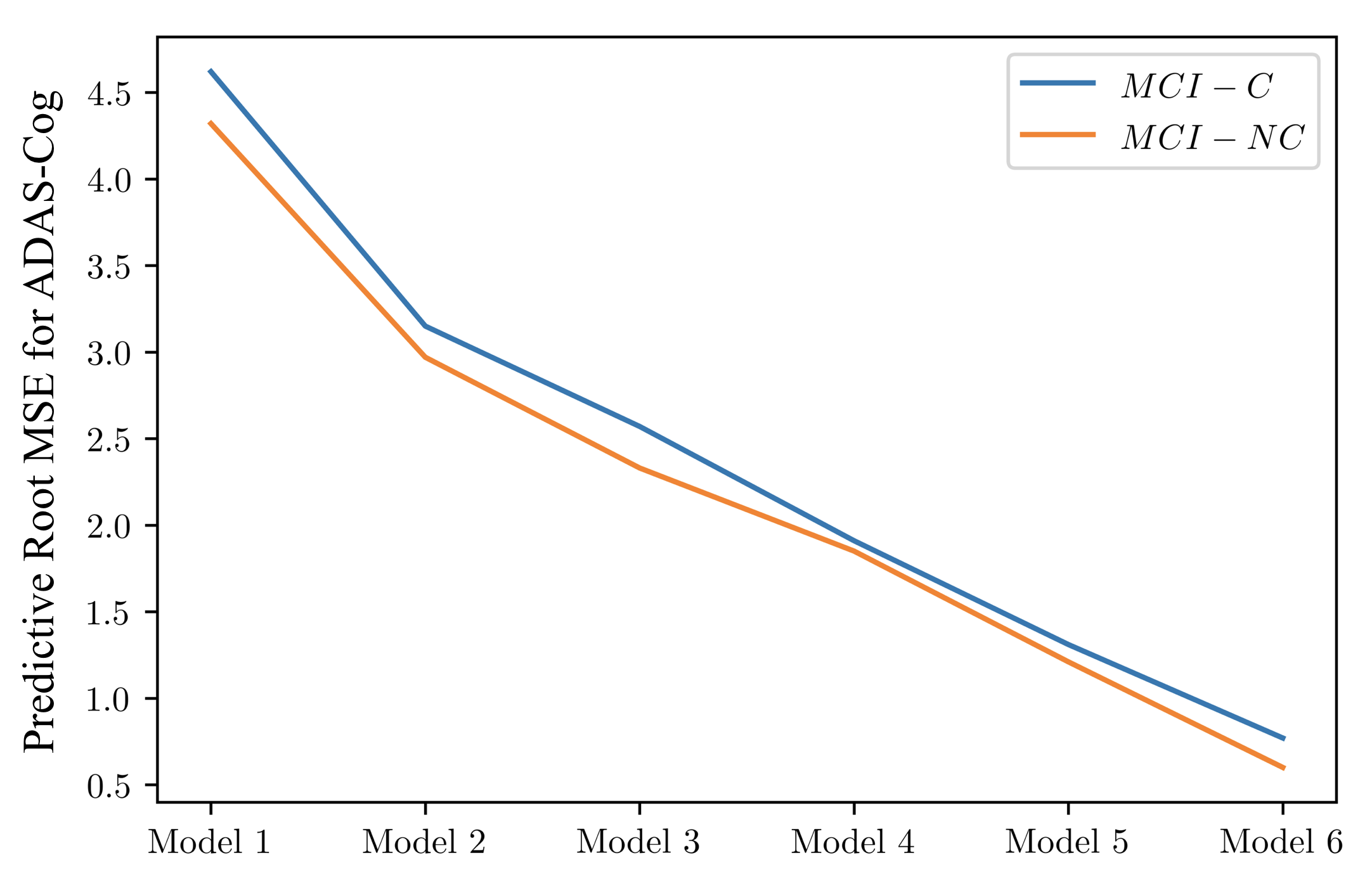}
\caption{The prediction comparisons of our method using six levels of longitudinal data. The predictive root MSEs of ADAS-Cog scores for both MCI-C and MCI-NC subjects at M48 are reported.}
\label{predictmy}
\end{figure*}

Following the flowchart in Figure \ref{framework}, we perform the feature selection method (\ref{eqn:highdimenscheme}) and prediction as discussed in Section \ref{sec:materialmethod} for each of the six models. Take the training and testing for Model 3 as an example. The MCI-C and MCI-NC subjects are trained and tested separately. 
First, we randomly leave out half  samples of both MCI-C and MCI-NC subjects, respectively, for testing in each experiment. 
For the training of Model 3, we choose subjects that have data at bl or M06 or M12 and  use the longitudinal data in (\ref{eqn:highdimenscheme}) and (\ref{eqn:equivform}) to select features and estimate the varying coefficients. Here, $n=172$, $p=327$ and $\max_{i} m_i=3$. Tuning parameters including the $\lambda$ in (\ref{eqn:highdimenscheme}) and the $\tau_1$ in (\ref{eqn:equivform}) are selected by
ten-fold cross-validation with the criteria that minimizes the predictive root MSE of ADAS-Cog scores  of subjects at bl, M06 and M12. 
With the selected features and estimated coefficients, we estimate the values of covariates at M48 as discussed in Section \ref{sec:materialmethod} and  predict the  ADAS-Cog scores at M48 for both MCI-C and MCI-NC subjects. 
The experiment of training and testing for Model 3 is replicated  100 times. We summarized the averaged predictive root MSE for ADAS-Cog scores at M48 in Figure \ref{predictmy}. 
Similarly, we train and test all  other five models besides Model 3. 
It is evident in Figure \ref{predictmy}  that the longitudinal information can significantly improve the prediction results compared with only using baseline information. And the more longitudinal data included, the better prediction results are obtained.
We also observe that prediction results for MCI-NC are better than those for MCI-C. This observation is expected since MCI-NC subjects have more stable clinical status and less varied clinical scores. 

\begin{figure*}[h!]
\centering
\vspace{0.1in}
\includegraphics[width=0.8\textwidth]{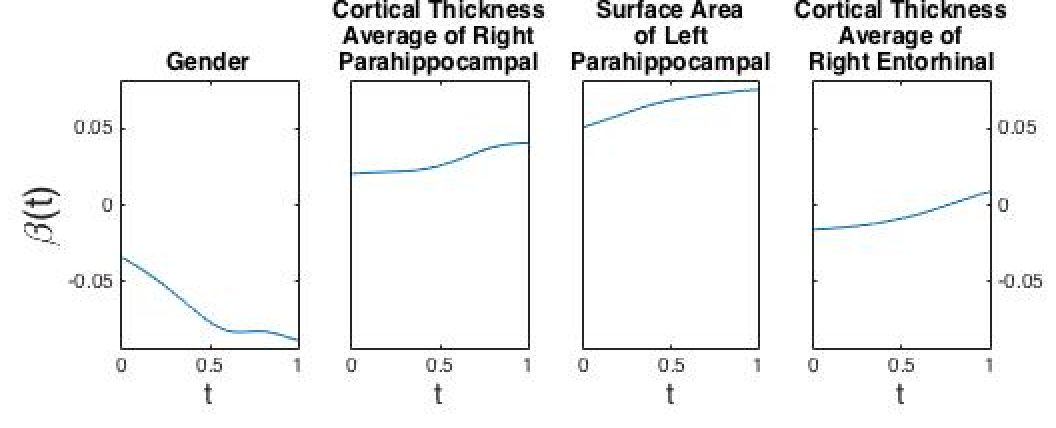}
\caption{Examples of selected features in training Model 6. The time $t$ is scaled to $[0,1]$, where $t=0$ represents bl and  $t=1$ represents M48.}
\label{fig:feature}
\end{figure*}

We show in Figure \ref{fig:feature} the four features that are consistently selected in 100 experiments in training Model 6. These learned features agree with the existing medical discoveries on AD-related features; See \cite{lin2015marked} for gender,  \cite{tognin2014reduced} for the cortical thickness of the right parahippocampal cortex, \cite{panizzon2009distinct} for the surface area of the left parahippocampal cortex, and \cite{velayudhan2013entorhinal} for the cortical thickness of the right entorhinal cortex.
Figure \ref{fig:feature} also illustrates the varying coefficients of these selected features. The maximum effect of each MRI ROI varies throughout disease progression, which suggests that different MRI ROIs have different functional relations with the clinical ADAS-Cog score. 
This observation confirms the evidence and hypothesis in \citet{sabuncu2011dynamics} and \citet{schuff2012nonlinear} that atrophy does not affect all regions of the brain simultaneously, but  in a sequential manner.

Now we compare our method (\ref{eqn:highdimenscheme})  with other two state-of-the-art methods:
\begin{itemize}
\item The longitudinal analysis in \cite{chincarini2016integrating}, which only uses the hippocampal volume shrinkage rate as the feature and assumes the longitudinal trend being a linear map. This method is different from our proposed method (\ref{eqn:highdimenscheme}) that uses all MRI ROIs, including the hippocampal volume shrinkage as features and assumes the nonlinear longitudinal trend.
\item The longitudinal analysis in \cite{zhang2012predicting}, which depends on linear feature representations and a group Lasso for variable selection  \citep{yuan2006model}.  This is different from our proposed  method (\ref{eqn:highdimenscheme}) that uses nonlinear feature representations in the varying coefficient model (\ref{eqn:varymodel}).
\end{itemize}
\begin{figure*}[h!]
\centering
\includegraphics[width=0.7\textwidth]{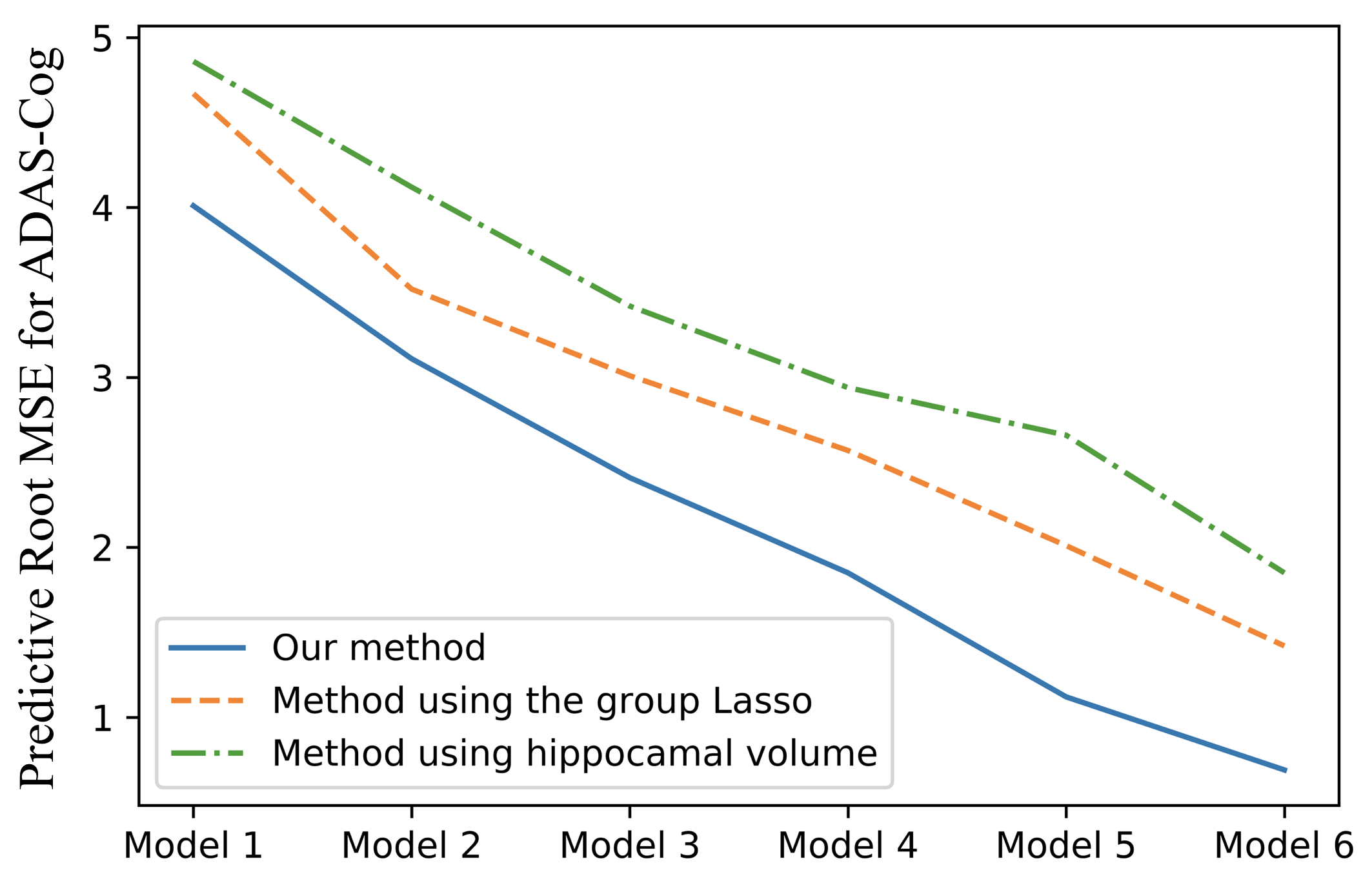}
\caption{The prediction comparisons of three methods  for MCI-C. The predictive root MSEs of ADAS-Cog scores for MCI-C subjects at M48 are reported.}
\label{predictmcic}
\end{figure*}

\begin{figure*}[h!]
\centering
\includegraphics[width=0.7\textwidth]{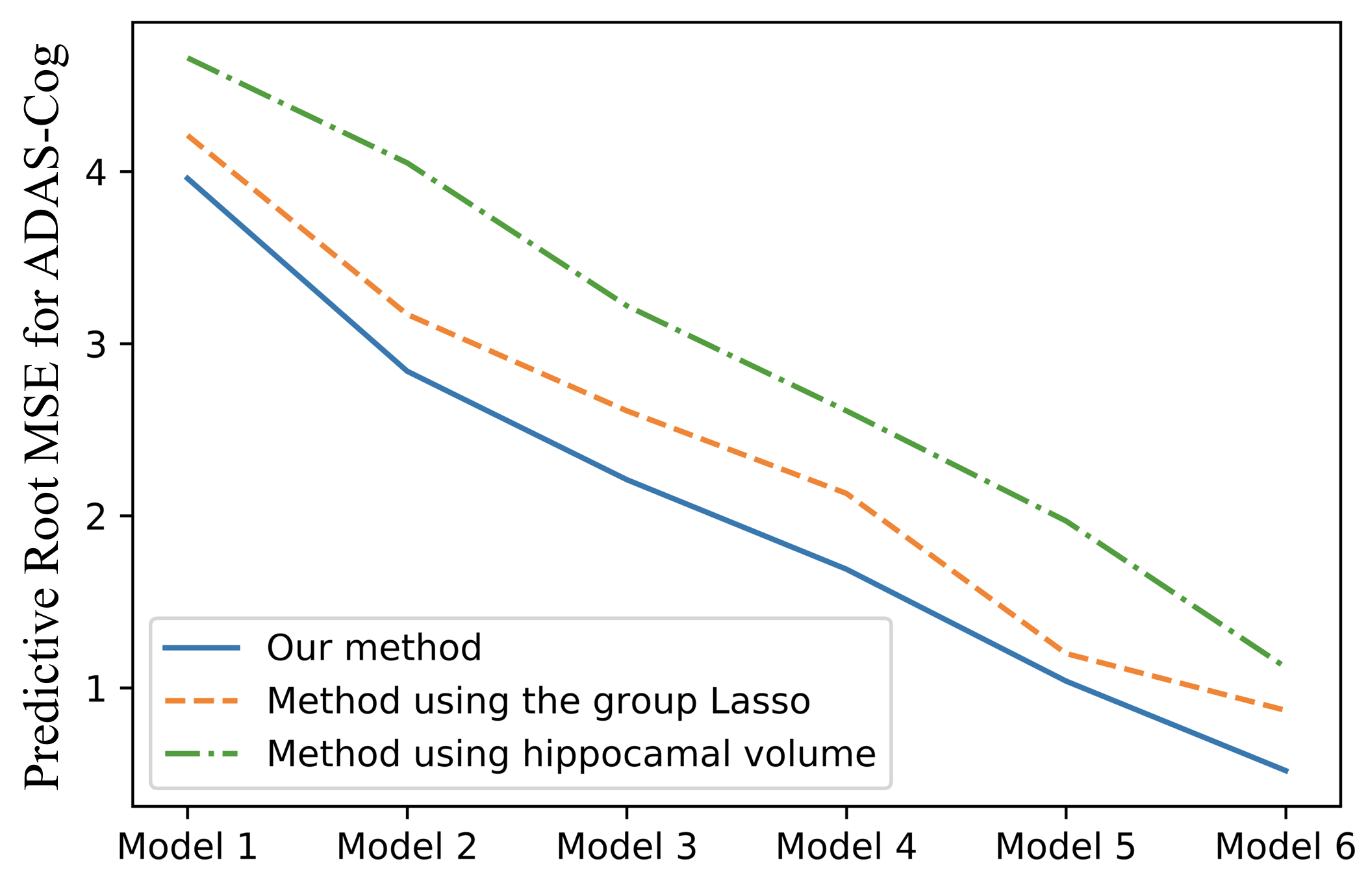}
\caption{The prediction comparisons of three methods  for MCI-NC. The predictive root MSEs of ADAS-Cog scores for  MCI-NC subjects at M48 are reported.
}
\label{predictmcinc}
\end{figure*}

Since the methods in \citet{chincarini2016integrating} and \citet{zhang2012predicting} require identical scanning times and an equal number of scans across samples, we perform Model 1--6 for AD prediction with data of subjects having no missing visits. 
In each experiment, we randomly leave out half of the samples in both MCI-C and MCI-NC for prediction. 
For the training of each model, we use the ten-fold cross-validation to select  tuning parameters in (\ref{eqn:highdimenscheme}) and also for the methods in \citet{chincarini2016integrating} and \citet{zhang2012predicting}.
We replicate each experiment for $100$ times. 
The prediction comparison results for MCI-C are shown in Figure  \ref{predictmcic} and the prediction comparison results for MCI-NC are summarized in Figure  \ref{predictmcinc}, where the predictive root MSEs for ADAS-Cog at M48 are reported. 
It is clear that our method consistently achieves better prediction performances for both MCI-C and MCI-NC. The reasons are that our method models the nonlinear progression of longitudinal features and it selects AD-related features from the whole brain MRI instead of  using  only prespecified features for prediction.

\section{Discussion}
\label{sec:discussion}
We study a framework to integrate longitudinal features from the structural MR images for AD prediction based on
varying coefficient models. 
We propose  a novel variable selection method by combining  smoothing splines and Lasso, which enables simultaneous selection and estimation and is adaptive to heterogeneous longitudinal data. 
Our work is the first in the literature to model nonlinear progressions of longitudinal features in the high-dimensional setting.
For validating our method, we conduct experiments with the ADNI dataset and show that the proposed method outperforms the state-of-the-art longitudinal analysis methods. 
 It is promising and easy to implement the proposed method in other longitudinal data analysis examples.

It would be interesting to use the predicted ADAS-Cog scores in a statistical classification model to predict whether a new MCI subject will become a converter or not. Here, the MCI-C and MCI-NC subjects need to be trained together, which is slightly different from experiments in Section \ref{sec:experiment}. Furthermore, we only use MR images for AD prediction in this paper. It would be of great interest to apply the proposed method to integrate multi-modal data, including MRI, PET, and functional MRI. We expect the integration of multi-modal information would further improve the accuracy of the AD prediction.
Another important future direction is to understand whether incorporating the known or estimated correlation structure between measurements at the same covariate,   $x_{ij}(t_{i1}),\ldots,x_{ij}(t_{im_i})$, into our method (\ref{eqn:highdimenscheme}) and (\ref{eqn:equivform}) can improve the selection or estimation results for varying coefficients $\beta_1(\cdot),\ldots,\beta_p(\cdot)$.

\bigskip
\section*{Appendix}
\appendix

\section{ADNI Database Description}
\label{sec:adni1}

The ADNI was launched in 2003 as a public-private
partnership, led by Principal Investigator Michael W. Weiner, MD. The primary goal of ADNI has been to
test whether serial magnetic resonance imaging (MRI), positron emission tomography (PET), other
biological markers, and clinical and neuropsychological assessment can be combined to measure the
progression of mild cognitive impairment (MCI) and early Alzheimer’s disease (AD).

Clinical diagnosed Alzheimer's disease patients   must have had mild AD and had to meet the National Institute of Neurological and Communicative Disorders and Stroke--Alzheimer's Disease and Related Disorders Association (NINCDS/ADRDA) criteria for probable AD in \cite{mckhann1984clinical}. The mild cognitive impairment subjects should have largely intact general cognition as well as  functional performance. Study subjects should have been given written informed consent at the time of enrollment for imaging and genetic sample collection and completed questionnaires approved by each participating sites Institutional Review Board (IRB).

\section{Preprocessing of the Brain MRI Used Here}
\label{sec:preprocessbrainmri}

The structural MRI used in this study are cortical gray matter volumes processed using FreeSurfer software version 4.4 longitudinal image processing framework (\url{https://surfer.nmr.mgh.harvard.edu/}) (``ucsffsl" file).
This dataset has been used in, e.g., \citet{tosun2011relationship}, \citet{toledo2014csf}, and \citet{dai2017high}.
Specifically, subjects with a 1.5-T MRI  were included in the dataset where the scans were preprocessed
by certain correction methods including gradwarp, B1 calibration, N3 correction, and skull-stripping (see, e.g., \cite{jack2008alzheimer} for detail), and the FreeSurfer 4.4 implements the symmetric registration \citep{reuter2010highly} and unbiased robust template estimation \citep{reuter2012within}. 
Only MRIs which passed the quality control for all the areas were included in our study.
There are total 393 ROIs of brain MRI created by FreeSurfer 4.4 and they consist of volumes of brain regions obtained after cortical parcellation and white matter parcellation, surface area of the brain regions and cortical thickness of the brain regions.
However, some ROIs are missing more than $90\%$ across all samples due to the preprocessing. In the Section \ref{sec:experiment}, we use 324 ROIs with at most $20\%$ missing values across the preprocessed samples. 


\section{Proof of Theorem \ref{thm:existminimizer}}
\label{proofthm:existminimizer}
Denote by $A(b,\beta_1(\cdot)\ldots,\beta_p(\cdot))$ the functional to be minimized in (\ref{eqn:highdimenscheme}).
It is clear that $A(b,\beta_1(\cdot)\ldots,\beta_p(\cdot))$ is convex and continuous in $\beta_j(\cdot)$s.
Denote by $J(\beta_1(\cdot)\ldots,\beta_p(\cdot)) = \lambda\sum_{j=1}^p\|\beta_j\|_{\HH_K}$, and
without loss of generality, we assume $\lambda=1$. 
Denote by $c_K=\max_{i, \nu}K^{1/2}(t_{i\nu},t_{i\nu})$ and  $c_x=\max_{j, i, \nu}|x_{ij}(t_{i\nu})|$.
By Cauchy-Schwarz inequality, for any $i=1,\ldots,n$, $\nu=1,\ldots,m_i$, 
\begin{equation}
\label{eqn:bndonbetajxijtinu}
\begin{aligned}
&\left\vert\sum_{j=1}^p\beta_j(t_{i\nu})x_{ij}(t_{i\nu})\right\vert  = \left\vert\left\langle\sum_{j=1}^p\beta_j(\cdot)x_{ij}(t_{i\nu}), K(t_{i\nu},\cdot)\right\rangle_{\HH_K}\right\vert\\
& \leq \left\|\sum_{j=1}^p\beta_j(\cdot)x_{ij}(t_{i\nu})\right\|_{\HH_K}K(t_{i\nu},t_{i\nu})\leq c_K \left\|\sum_{j=1}^p\beta_j(\cdot)x_{ij}(t_{i\nu})\right\|_{\HH_K}\leq c_Kc_xJ(b,\ldots,\beta_p).
\end{aligned}
\end{equation}
Denote $\rho=\max_{i,\nu}\{y_{i\nu}^2+|y_{i\nu}|+1\}$. Consider the set
\begin{equation*}
\begin{aligned}
\Omega & = \{\beta_1(\cdot),\ldots,\beta_p(\cdot)\in\HH_K, b\in\R: J(\beta_1(\cdot),\ldots,\beta_p(\cdot))\leq \rho, |b|\leq \rho^{1/2}+(c_Kc_x+1)\rho\}.
\end{aligned}
\end{equation*}
Since $\Omega$ is closed, convex, and bounded set, there exists a minimizer for  (\ref{eqn:highdimenscheme}) in $\Omega$. Denote the minimizer by $\tilde{\beta}_0, \tilde{\beta}_1(\cdot),\ldots,\tilde{\beta}_p(\cdot)$. Then, $A( \tilde{\beta}_0,\tilde{\beta}_1(\cdot),\ldots,\tilde{\beta}_p(\cdot))\leq A(0,0,\ldots,0)<\rho$.
On the other hand, for any $\beta_1(\cdot),\ldots,\beta_p(\cdot)\in\HH_K$ satisfying $J(\beta_1(\cdot),\ldots,\beta_p(\cdot))>\rho$. It is clear that $A(b,\beta_1(\cdot)\ldots,\beta_p(\cdot))\geq J(\beta_1(\cdot),\ldots,\beta_p(\cdot))>\rho$. For any $\beta_1(\cdot),\ldots,\beta_p(\cdot)\in\HH_K$ with $J(\beta_1(\cdot),\ldots,\beta_p(\cdot))\leq \rho$ and $|b|>\rho^{1/2}+(c_Kc_x+1)\rho$, (\ref{eqn:bndonbetajxijtinu}) implies that for any $i=1,\ldots,n$, $\nu=1,\ldots,m_i$, 
\begin{equation*}
\begin{aligned}
& \left|b+\sum_{j=1}^p\beta_j(t_{i\nu})x_{ij}(t_{i\nu}) - y_{i\nu}\right| > \rho^{1/2}+(c_Kc_x+1)\rho - c_Kc_x\rho-\rho=\rho^{1/2}.
\end{aligned}
\end{equation*}
Hence, $A(b, \beta_1(\cdot),\ldots,\beta_p(\cdot))>\rho$. Therefore, for any $b,\beta_1(\cdot),\ldots,\beta_p(\cdot)\not\in\Omega$, we have that $A(b,\beta_1(\cdot),\ldots,\beta_p(\cdot))>A(\tilde{\beta}_0,\tilde{\beta}_1(\cdot),\ldots,\tilde{\beta}_p(\cdot))$, where $\tilde{\beta}_0, \tilde{\beta}_1(\cdot),\ldots,\tilde{\beta}_p(\cdot)$ is the minimizer of (\ref{eqn:highdimenscheme}). We complete the proof.


\section{Algorithm for Solving the Minimizer of (\ref{eqn:highdimenscheme})}
\label{subsec:algorithm}


We provide an algorithm for solving the minimizer of (\ref{eqn:highdimenscheme}). The algorithm is based on  Theorem \ref{thm:equivalence}, whose proof will be given later  in Appendix \ref{proofofequivalence}.
Consider for any fixed $\theta_1,\ldots,\theta_p\geq 0$. If $\theta_j=0$ for some $j$, then $\beta_j=0$ in 
 the optimization (\ref{eqn:equivform}).
Without less of generality, let $\theta_1,\ldots,\theta_p>0$. Then (\ref{eqn:equivform})  is equivalent to the smoothing spline type problem: find $b\in\R,\beta_1(\cdot),\ldots,\beta_p(\cdot)\in\HH_K$ to minimize
\begin{equation}
\label{eqn:ssestimation}
 \begin{aligned}
 &  \frac{1}{N}\sum_{i=1}^n\sum_{\nu=1}^{m_i}[y_{i\nu} - b-\sum_{j=1}^p \beta_j(t_{i\nu})x_{ij}(t_{i\nu})]^2+ \sum_{j=1}^p (\tau_0\theta^{-1}_j)\|\beta_j\|_{\HH_K}^2.
   \end{aligned}
\end{equation}
 By the representer lemma \citep{wahba1990spline}, $\beta_1(\cdot),\ldots,\beta_p(\cdot)$ have a closed form expression: 
 \begin{equation*}
\beta_j(t) = \sum_{i=1}^n\sum_{\nu=1}^{m_i}c^j_{i\nu}K(t_{i\nu},t), \quad \forall j=1,\ldots,p.
\end{equation*}
Define a $m_{i_1}\times m_{i_2}$ matrix $\Sigma_j^{(i_1,i_2)}$ by
\begin{equation*}
\begin{aligned}
 \Sigma_j^{(i_1,i_2)} =   \left( \begin{array}{ccc}
x_{i_1j}(t_{i_11})K(t_{i_21}, t_{i_11}) & \cdots & x_{i_1j}(t_{i_11})K(t_{i_2m_{i_2}}, t_{i_11}) \\
\vdots&  & \vdots \\
x_{i_1j}(t_{i_1{m_{i_1}}})K(t_{i_21}, t_{i_1m_{i_1}}) & \cdots & x_{i_1j}(t_{i_1m_{i_1}})K(t_{i_2m_{i_2}}, t_{i_1m_{i_1}}) 
 \end{array} \right)
 \end{aligned}
\end{equation*}
and let $\Sigma_j$ be a $N\times N$ ($N=\sum_{i=1}^nm_i$) matrix where the $(i_1,i_2)$th $m_{i_1}\times m_{i_2}$ matrix is $\Sigma_j^{(i_1,i_2)}$.
Define kernel matrix $\Sigma$ by
\begin{equation*}
\Sigma = \left( \begin{array}{cccc}
\Sigma_1 & \Sigma_2 & \cdots & \Sigma_p 
\end{array} \right)\in\R^{N\times N\cdot p}.
\end{equation*}
Let the unknown coefficient vector $c^j$  be
\begin{equation*}
c^j = \left( \begin{array}{ccccccc}
c_{11}^j & \cdots & c_{1m_1}^j & \cdots  & c_{n1}^j & \cdots & c_{nm_n}^j
\end{array} \right)^\top\in\R^{N},
\end{equation*}
and 
\begin{equation*}
c = \left( \begin{array}{cccc}
\{c^1\}^{\top} &  \{c^2\}^{\top} & \cdots & \{c^p\}^{\top}
\end{array} \right)^\top\in\R^{Np}.
\end{equation*}
Write the response vector $y$ as
\begin{equation*}
y = \left( \begin{array}{ccccccccccccccc}
y_{11} & \cdots & y_{1m_1} & \cdots  & y_{n1} & \cdots & y_{nm_n} \end{array} \right)^\top\in\R^{N}.
\end{equation*}
Let $\mathbf{1}_N$ be the column vector consisting of $N$ $1$'s. Then (\ref{eqn:ssestimation}) becomes
\begin{equation*}
\begin{aligned}
& \frac{1}{N}\left(y - \Sigma c - b\mathbf{1}_{N}\right)^\top\left(y - \Sigma c - b\mathbf{1}_{N}\right)+\sum_{j=1}^p(\tau_0\theta^{-1}_j)\{c^j\}^\top \Sigma_j c^j,
\end{aligned}
\end{equation*}
which has the unique solution  given as follows:
\begin{equation}
\label{eqn:solbandc}
\begin{aligned}
\hat{b} & = [\mathbf{1}_N^\top(\mathbf{1}_{N\times N}-\Sigma\tilde{\Sigma}^{-1}\Sigma^\top)\mathbf{1}_{N}]^{-1}\cdot\mathbf{1}_N^\top(\mathbf{1}_{N\times N}-\Sigma\tilde{\Sigma}^{-1}\Sigma^\top)y,\\
\hat{c} & = \tilde{\Sigma}^{-1}\Sigma^\top(y-\mathbf{1}_{N}\hat{b}),
\end{aligned}
\end{equation}
where $\tilde{\Sigma} = \Sigma^\top\Sigma + N\text{diag}\{(\tau_0\theta^{-1}_1)\Sigma_1,\ldots,(\tau_0\theta^{-1}_p)\Sigma_p\}$.

Note that when $\theta_1,\ldots,\theta_p$ are fixed, (\ref{eqn:equivform}) is equivalent to find $b\in\R, c\in\R^{Np}$ to minimize
\begin{equation}
\label{eqn:fixedtheta}
\begin{aligned}
& \frac{1}{N}(y - b\mathbf{1}_{N} - \sum_{j=0}^p\theta_j\Sigma_j c^j)^\top\cdot(y  - b\mathbf{1}_{N} - \sum_{j=0}^p\theta_j\Sigma_j c^j)+\sum_{j=0}^p(\tau_0\theta_j)\{c^j\}^\top \Sigma_j c^j.
\end{aligned}
\end{equation}
The minimizer of (\ref{eqn:fixedtheta}) is
\begin{equation*}
b = \hat{b}\quad\text{and} \quad c^j = \theta_j^{-1}\hat{c}^j, \quad j=0,1,\ldots,p,
\end{equation*}
where $\hat{b}$ and $\hat{c}$ are given by (\ref{eqn:solbandc}).

On the other hand, consider when $c$ is fixed, then the minimization of (\ref{eqn:equivform})  is equivalent to
\begin{equation*}
\begin{aligned}
 \min_{\theta,b}\|y - \sum_{j=0}^p\theta_j\Sigma_j c^j - b\mathbf{1}_{N}\|^2 
&+ N\tau_0\sum_{j=0}^p\theta_j\{c^j\}^\top \Sigma_j c^j +N\tau_1\sum_{j=0}^p\theta_j,\\
& \quad\quad\quad\quad\text{s.t. } \theta_j\geq 0, j=0,1,\ldots,p,
\end{aligned}
\end{equation*}
which can be written as
\begin{equation}
\label{eqn:fixedcandbnew}
\begin{aligned}
 \min_{\theta,b}\|y - \sum_{j=0}^p\theta_j\Sigma_j c^j - b\mathbf{1}_{N}\|^2 
& + N\tau_0\sum_{j=0}^p\theta_j\{c^j\}^\top \Sigma_j c^j,\\
& \quad\quad\text{s.t. } \theta_j\geq 0, j=0,1,\ldots,p; \sum_{j=0}^p\theta_j\leq M,
\end{aligned}
\end{equation}
for some $M\geq 0$. 

Therefore, we propose an algorithm to iterate (\ref{eqn:fixedtheta}) and (\ref{eqn:fixedcandbnew}) to give the minimizer of  (\ref{eqn:equivform}). 
We observe in experiments that the objective function in optimization (\ref{eqn:equivform})   decreases quickly in the first iteration and after the first iteration the objective function is close to the objective function at convergence. 
It motivates us to consider the following one-step update algorithm:
\begin{itemize}
\item[1.] Initialization: fix $\theta_j=1$ for $j=0,1,\ldots,p$.
\item[2.] Solve for $c$ and $b$ in (\ref{eqn:fixedtheta}) and tune $\tau_0$
according to the generalized cross-validation (GCV). Fix $\tau_0$ at the chosen value in all later steps.
\item[3.] For $c$ and $b$ obtained in step 2, solve for $\theta$ in  (\ref{eqn:fixedcandbnew}) with a fixed $M$.
\item[4.] With $\theta$ obtained in step 3, solve for $c$ and $b$ in (\ref{eqn:fixedtheta}).
\end{itemize}
We choose the best $M$  in Step 3 according to the five-fold cross-validation in the experiments. 


\section{Proof of Theorem \ref{thm:equivalence}}
\label{proofofequivalence}
Recall that we introduce $A(b, \beta_1(\cdot),\ldots,\beta_p(\cdot))$ in Section \ref{proofthm:existminimizer} to denote the functional  to be minimized in (\ref{eqn:highdimenscheme}). Let $B(\theta_1,\ldots,\theta_p;b, \beta_1(\cdot),\ldots,\beta_p(\cdot))$ be the functional  in (\ref{eqn:equivform}). Note
\begin{equation*}
\begin{aligned}
\tau_0\theta_j^{-1}\|\beta_j\|_{\HH_K}^2+\tau_1\theta_j& \geq 2\tau_0^{1/2}\tau_1^{1/2}\|\beta_j\|_{\HH_K}  = \lambda^2\|\beta_j\|_{\HH_K}, \ \forall \theta_j\geq 0,
\end{aligned}
\end{equation*}
and the equality in the above formula holds if and only if $\theta_j = \tau_0^{1/2}\tau_1^{-1/2}\|\beta_j\|_{\HH_K}$. Therefore, 
\begin{equation*}
\begin{aligned}
& B(\theta_1,\ldots,\theta_p;b, \beta_1(\cdot),\ldots,\beta_p(\cdot)) \geq A(b, \beta_1(\cdot),\ldots,\beta_p(\cdot)), \  \ \forall \theta_j\geq 0,
\end{aligned}
\end{equation*}
and the equality holds if and only if $\theta_j = \tau_0^{1/2}\tau_1^{-1/2}\|\beta_j\|_{\HH_K}$ for all $j=1,\ldots,p$. We complete the proof.

\section*{Acknowledgements}

Data used in preparation of this article were obtained
from the Alzheimer's Disease Neuroimaging Initiative (ADNI) database (\url{www.adni.usc.edu}). As such, the investigators within the ADNI contributed to the design and implementation of ADNI and/or provided data but did not participate in analysis or writing of this report. A complete listing of ADNI investigators can be found at \url{http://adni.loni.usc.edu/wp-content/uploads/how_to_apply/ADNI_Acknowledgement_List.pdf}.

ADNI  (National Institutes of Health Grant U01 AG024904 and Department of Defense award
number W81XWH-12-2-0012) is funded by the National Institute on Aging, the National Institute of
Biomedical Imaging and Bioengineering, and through generous contributions from the following: AbbVie,
Alzheimer’s Association; Alzheimer’s Drug Discovery Foundation; Araclon Biotech; BioClinica, Inc.; Biogen;
Bristol-Myers Squibb Company; CereSpir, Inc.; Cogstate; Eisai Inc.; Elan Pharmaceuticals, Inc.; Eli Lilly and
Company; EuroImmun; F. Hoffmann-La Roche Ltd and its affiliated company Genentech, Inc.; Fujirebio; GE
Healthcare; IXICO Ltd.; Janssen Alzheimer Immunotherapy Research \& Development, LLC.; Johnson \&
Johnson Pharmaceutical Research \& Development LLC.; Lumosity; Lundbeck; Merck \& Co., Inc.; Meso
Scale Diagnostics, LLC.; NeuroRx Research; Neurotrack Technologies; Novartis Pharmaceuticals
Corporation; Pfizer Inc.; Piramal Imaging; Servier; Takeda Pharmaceutical Company; and Transition
Therapeutics. The Canadian Institutes of Health Research is providing funds to support ADNI clinical sites
in Canada. Private sector contributions are facilitated by the Foundation for the National Institutes of Health
(\url{www.fnih.org}). The grantee organization is the Northern California Institute for Research and Education,
and the study is coordinated by the Alzheimer’s Therapeutic Research Institute at the University of Southern
California. ADNI data are disseminated by the Laboratory for NeuroImaging at the University of Southern
California.

The author would like to thank the Editor,  the Associate Editor, and two referees for their helpful and constructive comments.

\section*{Funding}

The research was supported in part by NSF Grant DMS-1308877.

\bibliographystyle{Chicago}

\bibliography{references}

\end{document}